# SABER PRO SUCCESS PREDICTION MODEL USING DECISION TREE BASED LEARNING

Gregorio Pérez Bernal

Universidad EAFIT

Colombia

gperezb1@eafit.edu.co

Luisa Toro Villegas

Universidad EAFIT

Colombia

ltorov@eafit.edu.co

Mauricio Toro

Universidad EAFIT

Colombia

mtorobe@eafit.edu.co

**ABSTRACT**

The primary objective of this report is to determine what influences the success rates of students who have studied in Colombia, analyzing the Saber 11, some socioeconomical aspects and comparing the Saber Pro results with the national average. The problem this faces is to find what influences success, but it also provides an insight in the countries education dynamics and predicts one's opportunities to be prosperous. The opposite situation to the one presented in this paper could be the desertion levels, in the sense that by detecting what makes someone outstanding, these factors can say what makes one unsuccessful.

The solution proposed to solve this problem was to implement a CART decision tree algorithm that helps to predict the probability that a student has of scoring higher than the mean value, based on different socioeconomical and academic factors, such as the profession of the subject's parents and the results obtained on Saber 11. It was discovered that one of the most influential factors is the score in the Saber 11, on the topic of Social Studies, and that the gender of the subject is not as influential as it is usually portrayed as. The algorithm designed provided significant insight into which factors most affect any given person's success probability and if further pursued could be used in many given situations such as deciding which subject in school should be given more intensity to and academic curriculum in general.

## 1. INTRODUCTION

The Colombian education system is known to be a flawed one, having one of the highest superior education desertion rates in Latin America with a shocking 42% in 2018, according to el Banco Mundial (Casas Mogollón, 2018). One of the priorities the country ought to have is to determine the reasons why this is occurring and how to stop it, or at least to drastically reduce it. The present work will use the rising technology of machine learning, which is an artificial intelligence (AI) and as the name says it, it focuses on how a computer can learn from "experience", which in the context refers to data that had been previously classified and it uses information gain to learn the most of each factor.

## 2. PROBLEM

During this paper, the authors will attempt to predict the level of success of a given sample of people who have gone through the Colombian education system by exploring how certain aspects, such as their socio-economic situation, their highest education degree, their access to technology, and their Saber 11 results, which is the exam that high school graduates take in Colombia in their senior year. The success will be measured by whether said person obtained a higher than average score in the Saber Pro, which is the exam that every university student must take in order to graduate, exam or not.

The following project is of vast importance to Colombian society now a days due to the fact that in order to keep growing as a country, it needs to focus on the educating the youth, which represents the work force of our future, and Colombia has had a distinct history of only educating properly a selected few: the wealthy. This is not sustainable in the long run, therefore, it's primordial to determine and detect what is failing in the education system and correct it immediately.

In order to approach this problem, the authors will use an algorithm called decision trees, that is a type of machine learning, not unlike logistic regression, artificial neural networks and random forests.

## 3. RELATED WORK

The following section will briefly explain some types of decision trees and explore their advantages and disadvantages. All these algorithms have to decide how to act on certain aspects like which attributes to use for splitting, in which order to split them, number of splits, pruning (which data is important and which is not), and the stopping criteria. (Bhatt, Mehta, & D'mello, 2016)

Moreover, for each type, a real-life example of how the algorithms have been used shall be given.

Before commencing, a few terms shall be introduced.

- Entropy: in technology, Oxford defines entropy as "a way of measuring the lack of order that exists in a system." (Saha, 2018) In decision trees, entropy is a negative aspect, it decreases the certainty of the answer the algorithms provide us.
- Greedy splitting: "This is a numerical procedure where all the values are lined up and different split points are tried and tested using a cost function. The split with the best cost (lowest cost because we minimize cost) is selected" (Brownlee, 2016)
- Information gain: it "measures how much "information" a feature gives us about the class." (Sujan, 2018)
- Pruning: "Pruning refers to the removal of those branches in our decision tree which we feel do not contribute significantly to our decision process." (Saha, 2018)

### 3.1 ID 3 algorithm

The ID 3 (Iterative Dichotomiser 3) algorithm was first introduced by Ross Quinlan in 1975, which creates a decision tree from a given dataset. It offers the advantage that it is able to construct a small tree in a considerably short time. This tree is relatively quick due to the fact that while the data is being tested, the algorithm prunes it, reducing considerably the number of tests that are needed.

However, during the decision-making process, it only takes one attribute into account, which makes it hard to take into account problems of larger complexity. As well as that, implementing the ID3 algorithm with continuous variables may not work or may occupy a considerable amount of storage. This being said, the algorithm only works properly with discreet variables, which reduces the level of difficulty that this algorithm is able to solve.

The 1D3 algorithm was used in a study in 2015 in Mumbai University to predict campus placement, which refers to the hiring of college students before graduation, and this is evaluated based on their GPA, experience, programming skills and evaluation scores. The university used the study to help the performance of students in the future. (Bhatt, Mehta, & D'mello, 2016)

### 3.2 C4.5 algorithm

The C4.5 algorithm is a type of decision tree that was developed by Ross Quinlan, a pioneer in data mining. C4.5 is the successor of the ID 3 algorithm, and it brought two main improvements to the complexity of the problems that decision trees are able to solve. First, it includes the possibility of implementing discreet variables, which shaped the way in which problems were solved, for numeric values became available to be used. Moreover, this tree allows the existence of missing data to be implemented, meaning that whenever the value of something is unknown, the algorithm still works, guiding itself through the other bits of information presented.

On the other hand, C4.5 resolved the problem that the tree allows the existence of missing data to be implemented, however it backfired. Whilst this issue was cleared, as there are unknown variables that while the tree is being built, it constructs empty branches. Nodes with zero values or close to zero values have been encountered in multiple occasions. These values neither contribute to generate rules nor help to construct any class for classification task, thus, it only contributes by wasting valuable memory.

The C4.5 algorithm was used in a study about pregnancy risks and complications performed in 2016 by a group of professors which were published on the International Conference on Emerging Trends in Engineering, Science and Technology. The goal of the study was to standardize the information inquired during medical visits of women during pregnancy, giving priority to those who provide more information gain to the decision trees. It was shown by the algorithm that the most important factors were state, blood pressure, blood glucose level, weight, trimester and month. Then, they tried the algorithm on both standardized data and unstandardized data and showed that the prediction was more accurate for the first. (Lakshmi, Indumathi, & Ravi, 2016)

### 3.3 C 5.0 algorithm

The C5.0 algorithm was developed by Ross Quinlan, which was created to improve upon C4.5. When applied, it produces either a binary tree, that in which each node splits into two possible options, or multibranched tree, in which each node can split into any number of options. To determine into which factors it splits, it uses a method of measuring entropy which is information gain, and the goal is to maximize it, that way it takes better informed decisions. When presented with missing values or data, it roughly calculates its value basing itself upon the data of other variables.

This algorithm can be used to both create decision trees and rule-based models. Rule based models can use the structure of the tree to create mutually exclusive rules, it "assumes that the world can be understood as systems with

interrelated parts that perform predictable behaviors." (Whitney, 2017)

An application to the C 5.0 algorithm is for mapping the locations of possible groundwater. The investigation was done in Iran and it is of special importance due to the fact that said country is extremely dry and water is needed in almost every aspect of daily life. A dataset of already known spring locations was inputted, and some of their variables like "altitude, slope aspect, slope angle, plan curvature, profile curvature, topographic wetness index (TWI), slope length, distance from rivers and faults, rivers and faults density, land use, and lithology" (Golkarian, Naghibi, Kalantar, & Pradhan, 2018) were used.

### 3.4 CART (Classification and Regression Trees)

The CART algorithm was first described by Leo Breiman. It is characterized by the fact that it constructs binary trees, namely each internal node has exactly two outgoing edges.

CART enables the usage of both numerical and categorical variables, which places an advantage towards ID3. As well as that, it uses greedy splitting to select the input variables that will be used. Something very beneficial about CART is the fact that it is able to determine which variables are the most useful, and immediately prioritize them, and the most useless and it will automatically delete them.

The problem with CART comes with the fact that the tree that is generated is highly unstable. A slight modification of the learning sample may change considerably the way in which the results are obtained. Besides, CART splits only by one variable, which reduces the number of problems that this tree can manage to solve.

A study conducted in Humboldt University in Berlin found the many Financial Applications that the CART decision tree can have. It was done by simulating a market and results show how a CART-based business application can perform on stock market as well as utilizing CART as a forecasting system. The conclusion reached was that simulated financial performance was above the average level when CART was applied to the Simulation. This means that by implementing the CART algorithm, one can actually perform better in the stock market and eventually becoming wealthier.

(Andriyashin, 2005)

### 4. An array of arrays (Matrix)

In order to solve the problem, the data structure that was implemented is an array of arrays (shown in Figure 1), in which each element of the bigger array corresponds to a person, and each element of the person's array corresponds to a characteristic of them, for instance, one's ICFES exam score, or city of residency. It is important to point out that the first array of the large array corresponds to the headers of each column.

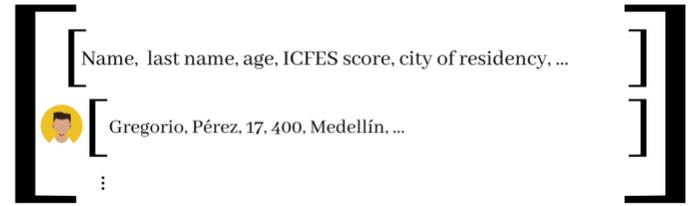

**Figure 1:** Matrix of persons. Each person is an individual array that contains a name, last name, age, ICFES score, city of residency and many other factors.

### 4.1 Operations of the data structure

For the data structure to solve the problem efficiently, the code will need certain operations to process the data given. The first and most crucial is the create operation, which will create a matrix of n*m filled with zero's (shown in Figure 2). The second is divided into two, access a specific element of a given person (Figure 3) and access a person in general (Figure 4). And the third operation is search (Figure 5), will when a certain factor is needed (e.g. name) it will search through the array [0], which is the array that will contain the column names and it will return the column number. The last operation is add (Figure 6) which replaces a row of zeros with a row with information, this in order to be more efficient than the append operation.

**Figure 2:** Create operation of the matrix.

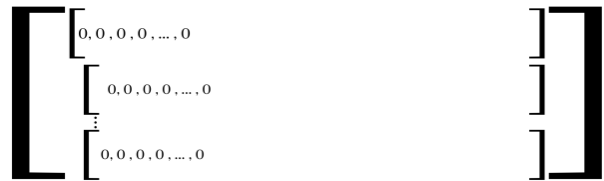

**Figure 3:** Access element operation of the matrix.

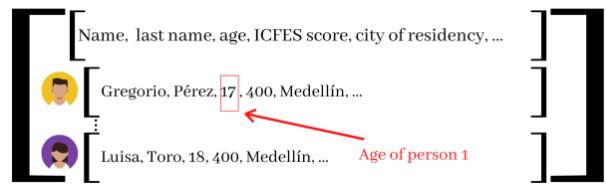

**Figure 4:** Access person operation of the matrix.

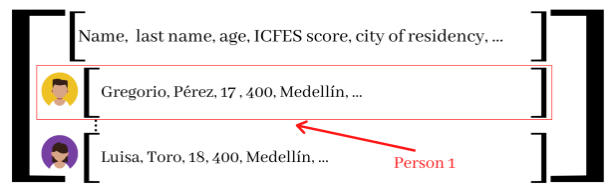

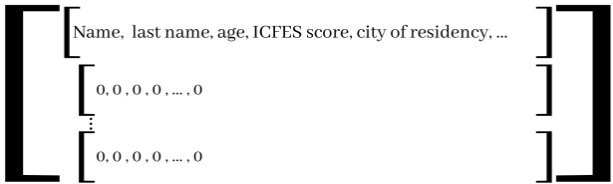

**Figure 5:** Add operation of the matrix.

### 4.2 Design criteria of the data structure

In order to have an efficient algorithm, the main data structure used will be one in the form of a matrix. This was chosen because of two fundamental reasons. First, the complexity of accessing any element of it is O(1), which is really beneficial. Another benefit is that accessing to information is extremely easy, due to the fact that everything is saved on a fixed position. The matrices are created with fixed position to avoid a flaw in python, the append operation which is inefficient.

### 4.3 Complexity analysis

The main reason why this data structure was selected is the fact that the complexity to access a single person or a characteristic of a person was found to have a complexity of O(1). The create matrix method was found to be O(m*n), n being the number of people and m being the number of characteristics per person, but it only does it once, avoiding a complexity of O(m*n^2) with append. As well as that, the method for deletion will not be considered, for the algorithm will examine which variables are useless and will ignore it instead of deleting them.

In Table 1, these complexities will be displayed according to each operation.

| Operations | Complexity of an array of arrays |
|---|---|
| Access | O(1) |
| Create | O(n*m) |

**Table 1:** Table to report complexity analysis (m refers to the number of people, n refers to the number of characteristics per person).

### 4.4 Execution time

In this section, the times taken to execute each operation according to each dataset are measured and are shown in Table 2.

| Method | Data set 1 (15,000) | Data set 2 (45,000) | Data set 3 (75,000) | Data set 4 (105,000) | Data set 5 (135,000) |
|---|---|---|---|---|---|
| Create | 0.027s | 0.031s | 0.031s | 0.034s | 0.032s |
| Access | 0.00s | 0.00s | 0.00s | 0.00s | 0.00s |

**Table 2:** Execution time of the operations of the data structure for each data set (worst case scenarios).

### 4.5 Memory used

In this section, the memory consumption of the first data structure implemented, the array of arrays or matrix, will be measured and shown according to which data set was used (it also specifies the amount of lines in each data set). The data obtained is shown in Table 3.

| | Data set 1 (15,000) | Data set 2 (45,000) | Data set 3 (75,000) | Data set 4 (105,000) | Data set 5 (135,000) |
|---|---|---|---|---|---|
| Memory | 6.344 KB | 6.346KB | 6.347 KB | 6.769KB | 6.771 KB |

**Table 3:** Memory used for each operation of the matrix data structure and for each data set data sets.

### 4.6 Result analysis

The results obtained in the previous section are optimal. The only considerable time-consuming operation is the adding operation, also called create operation, which is the one that inputs the data into the program, and considering the amount of data inputted, it's still good timing.

In comparison to other data structures, if queues, linked lists or stacks were used, in order to access to information, the complexity would be O(n), which presents a big disadvantage over arrays.

In the next table, the times measured will be compared with the times that other data structures would take.

| | Access | Search | Insert | Delete |
|---|---|---|---|---|
| Array | O(1) | O(n) | O(n) | O(n) |
| Linked List () | O(n) | O(n) | O(1) | O(1) |
| Stack | O(n) | O(n) | O(1) | O(n) |
| Queue | O(n) | O(n) | O(n) | O(1) |

**Table 4:** Analysis of the results

## 5. Classification and Regression Tree

In order to solve the problem, a decision tree that follows the guidelines of CART's algorithm was implemented. Basically, it reads data from results of previous years and analyses it, building a tree that shows which variables are the most important for a person to succeed. It builds the tree based on questions or conditions, for example whether the score in Saber 11 in social studies was greater than 52 or not, separating the data into two. In order to make it more friendly for the user, at the end of each branch of the tree a leaf will appear, in which it will calculate the percentage of probability of whether someone will be successful. As well as that, it will print the Gini impurity for each leaf, and use colors to represent furtherly the success of a person.

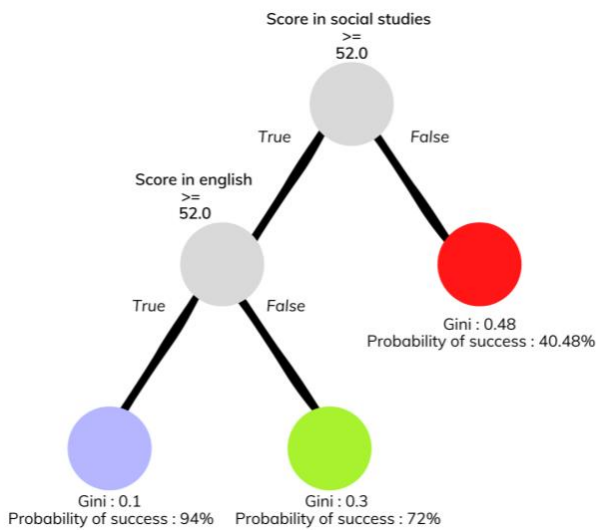

**Figure 6:** Decision tree.

### 5.1 Operations of the data structure

There are two main methods that the decision tree uses: create or train tree and classify new data using that tree. The former is that in order to create the tree, the variables are first separated as numerical or categorical. This is done because normally, numeric values are continuous whereas categorical ones are discrete, which means that they must be treated differently in order to have a more efficient algorithm. An example of how these variables are treated is to decide whether a student is successful regarding their score on the part of the test that focuses on math and the level of education of a parent. In order to build the tree for the level of education of the parent, one must ask, if the parent has either no education, secondary education, university level education and so on. Now, in order to measure the score on math, one cannot ask if the score was equal to 0.01 or 0.02 or 0.03 because it will be highly inefficient, so the operation of "greater than," in which the question asked would be If the score > 50, and then if score < 30 and so on.

Now, a function calculates the Gini impurity and the information gain of each variable and basing themselves on that, it finds which variables are the ones that not only are more important in order to succeed but also the ones that give more information on the process. In order to do that, a Python dictionary that serves to count how many cases of success and failures there are according to a question is implemented. After that, given how a question behaves, a partition is done, which splits the matrix into a true matrix and a false one. When this is done, all that is left to do it to create a tree. It is done recursively, testing all the possible partitions by every possible question and decides the one that has more information gain and constructs the tree basing itself on that. When the tree is all done, it looks like the tree displayed in figure 8 and it is ready to be tested on new data.

Then the latter, classifying new data, is basically using the conditions in the built tree to organize new data into it, taking a look person by person or record by record and going through the tree. Then, the records are all in a leaf, or node without further children or nodes coming out of it. Each leaf contains its success probability and if its greater than 50% then the person is predicted to be successful, and if not, the person is predicted to be unsuccessful in the Saber Pro.

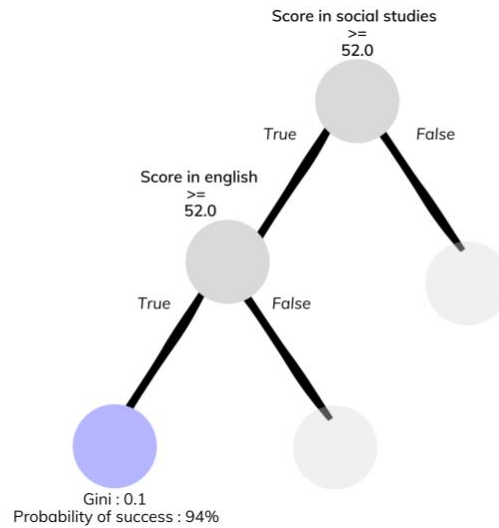

**Figure 7:** Training or creating the tree.

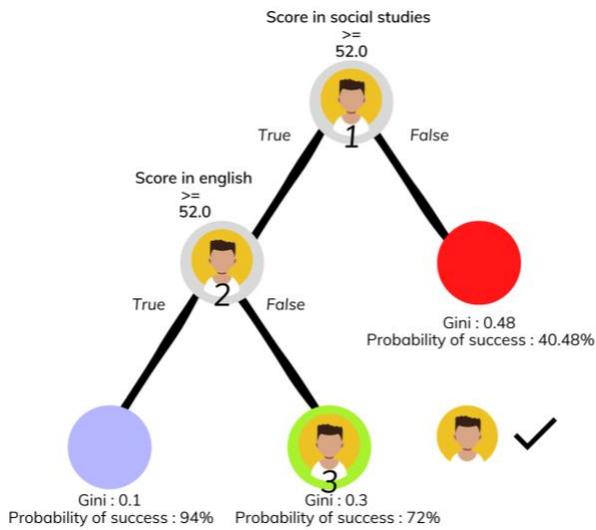

**Figure 8:** Classify or testing new data with the tree.

### 5.2 Design criteria of the data structure

The criteria used in order to create the data was regarding to have a more efficient code, specially timewise, because the program should be able to analyze data from more than 100,000 students. CART algorithm brought many advantages to the situation presented and the dataset used, for example that it handles well missing values, and that it can use categorical and numerical values. The algorithm identifies by itself which criteria are important, aiding thus with moral problems such as discrimination based on gender and ethnicity and by making a truer prediction based in the dataset. Lastly, with such a big dataset, it's important to be able to control the size of the tree and the CART algorithm allows it by limiting the recursion depth.

### 5.3 Complexity analysis

Derive the complexity of each operation of the data structure for the worst case and best case, As an example, this is a way to report the complexity analysis:

| Method | Complexity |
|---|---|
| Train tree | O(2^m) |
| Test new data | O(n*m) |

**Table 5:** Table to report complexity analysis

### 5.4 Execution time

In this section, the times taken to execute each operation according to each recursion depth with the dataset 5 (135,000 rows), are measured and are shown in Table 6.

| Method | Recursion depth 3 | Recursion depth 5 | Recursion depth 7 | Recursion depth 9 | Recursion depth 11 |
|---|---|---|---|---|---|
| Training | 160s | 270s | 316s | 433 s | 689s |
| Testing | 0.5s | 0.9s | 1.7s | 2.4 s | 2.5s |

**Table 6:** Execution time of the operations of the data structure for 135,000 records according to recursion depth.

### 5.5 Memory used

In this section, the memory consumption of the second data structure implemented, the decision tree or classification and regression tree, will be measured and shown according to which data set was used (it also specifies the amount of lines in each data set). The data obtained is shown in Table 7.

**Table 7:** Memory used for each operation of the data structure and for each data set data sets.

|  | Data set 1 (15,000) | Data set 2 (45,000) | Data set 3 (75,000) | Data set 4 (105,000) | Data set 5 (135,000) |
|---|---|---|---|---|---|
| Memory | 6.344 KB | 6.346KB | 6.347 KB | 6.769KB | 6.771 KB |

### 6. CONCLUSION

Regarding the present day problematic that the Colombian education system presents, it is crucial to find measures that give insight into which areas of knowledge are lacking and which people have more advantages and disadvantages than others according to some socioeconomical factors that should be breached. Present day technology has given the Colombian society many tools to help tackle this type of problems and in this specific situation one of the most prominent is machine learning using decision trees. This type of binary trees come in many different algorithms, which are explored in the present paper, but the one that showed most advantages was the Classification and Regression Tree Algorithm, also known as CART.

The algorithm used presents insight on the variables that most affect how one will do in the Saber Pro test. It was found that the score on the Social Studies section of the test is the most important to determine success. Score in Chemistry, Biology and whether the school setting was urban were also variables that highly affect the results. Furthermore, algorithm shows an accuracy percentage of around 80% depending on the size of the data set used to train it and the recursion depth, which is optimal and time efficient at around ten recursive calls.

Lastly, it is important to look beyond the algorithm and the mathematical terms and understand what each concept mean in real life. This algorithm is highly applicable in academic areas such as decision making for scholarships, according to how probable it is for the subject to succeed in Saber Pro and in regulating the admissions in certain universities. Moreover, the Colombian government could also benefit from it by using it to decide which areas of knowledge to fund or which personal factors to strengthen.

**6.1 Future work**

With every work, there always seems to be something left to do. In the case of the work presented in this paper, it would have been wonderful not only to create a decision tree, but the whole random forest, improving thus the prediction accuracy. Moreover, the algorithm used evaluates each criterion on its own, and sometimes misses some factors only the human eye could see, especially due to the way the dataset is presented. For example, some categorical criteria could be evaluated as a numerical value (greater than and lesser than, rather than equals), like the case of socioeconomical status. In terms of memory consumption, the data presented is in Spanish, and the algorithm treats the value as Unicode, when it could treat it as asci, saving half the memory consumption for the data structures. Lastly, the algorithm uses the data itself to build the tree, and that sometimes takes a toll on time efficiency, which could be solved by using an additional data structure to store the original data and assign it a key of sorts, which the algorithm could use to build the tree and then translate it into the original data.

**ACKNOWLEDGEMENTS**

Acknowledgements to EAFIT university for providing the course in which this project was done. As well as that, a thank you to Andrea Carvajal Maldonado and Juliana Restrepo Tobar which inspired part of the tree that was built.

**REFERENCES**

Andriyashin, A. 2005. *Financial Applications of Classification and Regression Trees*. Humboldt University, Berlin.

Bhatt, H., Metha, S. and D'mello, L. 2016. *Use of ID3 Decision Tree Algorithm for Placement Prediction*. International Journal of Computer Science and Information Technologies, Mumbai.

Brownlee, J. 2020. Classification And Regression Trees for Machine Learning. *Machine Learning Mastery*. https://machinelearningmastery.com/classification-and-regression-trees-for-machine-learning/.

Casas Mogollón, P. 2018. El problema no es solo plata: 42 % de los universitarios deserta | ELESPECTADOR.COM. *ELESPECTADOR.COM*. https://www.elespectador.com/noticias/educacion/el-problema-no-es-solo-plata-42-de-los-universitarios-deserta-articulo-827739.

GmbH, R. 2020. ID3 - RapidMiner Documentation. Docs.rapidminer.com. https://docs.rapidminer.com/latest/studio/operators/modeling/predictive/trees/id3.html.

Golkarian, A., Naghibi, S., Kalantar, B. and Pradhan, B. 2018. *Groundwater potential mapping using C5.0, random forest, and multivariate adaptive regression spline models in GIS*. Springer International Publishing.

IBM Knowledge Center. 2020. *Ibm.com*. https://www.ibm.com/support/knowledgecenter/en/SS3RA7_15.0.0/com.ibm.spss.modeler.help/c50node_general.htm.

Khadka, R. 2017. Machine Learning Types #2. *Medium*. https://towardsdatascience.com/machine-learning-types-2-c1291d4f04b1.

Kuhn, M. and Quilan, R. C5.0 Classification Models. Topepo.github.io. https://topepo.github.io/C5.0/articles/C5.0.html.

Lakshimi, B., Indumathi, T. and Ravi, N. 2016. *A study on C4.5 Decision Tree Classification Algorithm for Risk Predictions during Pregnancy*. International Conference on Emerging Trends in Engineering, Science and Technology, Karnataka, India.

Li, L. 2019. Classification and Regression Analysis with Decision Trees. *Medium*. https://towardsdatascience.com/https-medium-com-lorrli-classification-and-regression-analysis-with-decision-trees-c43cdbc58054.

Nguyen, A. Comparative Study of C5.0 and CART algorithms.

Rizvi, A. 2010. ID3 Algorithm. Sacramento, United States of America.

Saha, S. 2018. What is the C4.5 algorithm and how does it work?. *Medium*. https://towardsdatascience.com/what-is-the-c4-5-algorithm-and-how-does-it-work-2b971a9e7db0.

Sujan, N. 2018. What is Entropy and why Information gain matter in Decision Trees?. *Medium*. https://medium.com/coinmonks/what-is-entropy-and-why-information-gain-is-matter-4e85d46d2f01.

Tree algorithms: ID3, C4.5, C5.0 and CART. 2019. *Medium*. https://medium.com/datadriveninvestor/tree-algorithms-id3-c4-5-c5-0-and-cart-413387342164.

Whitney, V. 2017. Rule Based Modeling. *Medium*. https://medium.com/data-mining-the-city/rule-based-modeling-203b8af9fbc7